\documentclass[10pt,aps,twocolumn,floatfix,showpacs,showkeys,preprintnumbers,nofootinbib,superscriptaddress,longbibliography,bibnotes]{revtex4-1}
\pdfoutput=1
\usepackage[colorlinks=true,breaklinks=true]{hyperref}
\usepackage[utf8]{inputenc}
\hypersetup{allcolors=[rgb]{0.0 0.0 0.6},linkcolor=[rgb]{0.75 0.05 0.05}}
\usepackage{amsmath,amssymb}
\usepackage{epsfig}  
\usepackage{graphicx}   
\usepackage{grffile}
\usepackage{slashed}             
\usepackage{url}
\usepackage{color}
\usepackage{multirow}
\usepackage{placeins}
\usepackage[dvipsnames]{xcolor}
\usepackage{letltxmacro}
\LetLtxMacro{\oldcite}{\cite}
\renewcommand{\cite}[1]{\mbox{\oldcite{#1}}}
\usepackage[normalem]{ulem}

\allowdisplaybreaks

\bibpunct{[}{]}{,}{n}{}{,}

\begin{document}

\preprint{MPP-2020-226}

\title{Fast neutrino flavor  conversions  in one-dimensional core-collapse supernova models with and without muon creation}

\author{Francesco Capozzi}
\email{capozzi@vt.edu}  
\affiliation{Center for Neutrino Physics, Department of Physics, Virginia Tech, Blacksburg, VA 24061, USA
}
\affiliation{Max-Planck-Institut f\"{u}r Physik (Werner-Heisenberg-Institut), F\"{o}hringer Ring 6, 80805 M\"{u}nchen, Germany
}                     
\author{Sajad Abbar}
\email{abbar@mppmu.mpg.de}
\affiliation{Max-Planck-Institut f\"{u}r Physik (Werner-Heisenberg-Institut), F\"{o}hringer Ring 6, 80805 M\"{u}nchen, Germany
}      
\author{Robert~Bollig}
\email{bollig@mpa-garching.mpg.de}
\affiliation{Max-Planck-Institut f\"ur Astrophysik, Karl-Schwarzschild-Stra{\ss}e 1, D-85748 Garching, Germany}                            
\author{H.-Thomas~Janka}
\email{thj@mpa-garching.mpg.de}
\affiliation{Max-Planck-Institut f\"ur Astrophysik, Karl-Schwarzschild-Stra{\ss}e 1, D-85748 Garching, Germany}


\begin{abstract}
In very dense environments, neutrinos can undergo fast flavor conversions on  scales as short as a few centimeters provided that the angular distribution of the neutrino lepton number crosses zero. This work presents the first attempt to establish whether the non-negligible  abundance of muons and their interactions with neutrinos in the core of supernovae can affect the occurrence of such crossings. For this purpose we employ state-of-the-art one-dimensional core-collapse supernova simulations, considering models that include muon-neutrino interactions as well as models without these reactions. Although a consistent treatment of muons in the equation of state and neutrino transport  does not seem to modify significantly the conditions for the occurrence of fast modes, it  allows for the existence of an interesting phenomenon, namely fast instabilities in the $\mu-\tau$ sector. We also show that crossings  below the supernova shock are a relatively generic feature of the one-dimensional simulations under investigation, which contrasts with the previous reports in the literature. Our results highlight the importance of multi-dimensional simulations with muon creation, where our results must be tested in the future.
\end{abstract}

\maketitle

\section{Introduction}

At the end of their lives, massive stars ($M\gtrsim8M_\odot$) may undergo a violent explosion, labeled as core-collapse supernova (CCSN). During such an event, about $10^{53}$ ergs of energy is released into neutrinos of all flavors, corresponding to about 99\% of the total released gravitational energy. Such neutrinos are thought to be a crucial ingredient of the astrophysical processes leading to the final explosion.  Indeed, neutrinos might deposit enough energy to revive the shock, thus leading to the explosion. This is also known as the neutrino heating mechanism \cite{Janka:2017vcp,Janka:2017vlw,Burrows:2020qrp}. Neutrino interactions also affect  the electron fraction $Y_e$, which may indicate the presence of favorable conditions for the nucleosynthesis of heavy nuclei through r-processes, if its value is less than 0.5. Consequently, SN neutrinos represent an invaluable source of information of the  processes occurring in the exploding star.
  
   Neutrinos can experience flavor conversions during their propagation in the SN environment. In particular, in the highest density SN regions the neutrino self-interaction potential becomes large enough to induce the so-called collective conversions \cite{Duan:2010bg,Mirizzi:2015eza,Chakraborty:2016yeg,Tamborra:2020cul}, where neutrinos transform their flavors in a coherent fashion, despite having different oscillation frequencies  in vacuum. Such phenomena have a deeply non-linear nature and currently are not completely understood. This non linear evolution represents one of the most important unsolved problems in neutrino physics to date, since it impacts what we can learn from the next SN neutrino burst.

Lately, a lot of attention has been devoted to the so called fast flavor conversions \cite{Dasgupta:2017oko,Capozzi:2017gqd,Dasgupta:2018ulw,Capozzi:2018clo,Airen:2018nvp,Yi:2019hrp,Shalgar:2019kzy,Martin:2019gxb,Shalgar:2019qwg,Capozzi:2019lso,Johns:2019izj,Chakraborty:2019wxe,Shalgar:2020xns,Abbar:2017pkh,Capozzi:2020kge,Bhattacharyya:2020jpj,Johns:2020qsk,Shalgar:2020wcx}, which are triggered by a crossing between the angular distributions of neutrinos and antineutrinos, i.e., a crossing in the angular distribution of the neutrino electron lepton number (ELN). In other words, such crossings occur when there is an excess of neutrinos in a given range of propagation directions, whereas antineutrinos dominate in another one. If such a condition is satisfied just above the neutrino trapping region, fast conversions will develop on time scales as short as a few nanoseconds and might lead to a significant modification of the flavor content of the neutrino fluxes. Recent studies have looked for crossings in a plethora of 1D and multi-dimensional numerical simulations outputs \cite{Azari:2019jvr,Abbar:2018shq,DelfanAzari:2019tez,Abbar:2019zoq,Glas:2019ijo} and found them in the following regions: inside the proto-neutron star (PNS)~\cite{Abbar:2019zoq,Glas:2019ijo,DelfanAzari:2019tez}, in the neutrino decoupling region \cite{Abbar:2018shq} and in the pre-shock region \cite{Morinaga:2019wsv}. Each of these  zones rely on different microphysics for crossing generations.

The previous findings are all based on simulations adopting the standard assumption that the number densities and energy spectra of $\nu_\mu,\bar{\nu}_\mu,\nu_\tau,\bar{\nu}_\tau$ are equal. This is an effective approach where only three different neutrino species are simulated: $\nu_e$, $\bar{\nu}_e$ and $\nu_x$, where $\nu_x=\nu_\mu=\nu_\tau=\bar{\nu}_\mu=\bar{\nu}_\tau$. This assumption may only be taken  when a negligible muon content is assumed in the core of a SN, i.e. if one does not consider charged-current interactions of neutrinos  with muons. However, deep inside the supernova the electron chemical potential, the difference between the chemical potentials of neutrons and protons, and the temperature are of the order of several 10 MeV and higher. Under these conditions charged-current interactions of neutrinos with muons can happen~\cite{Fischer:2015lg,Bollig:2017lki,Bollig:2020,Fischer:2020vie}. When this is taken into account, one finds that both $\nu_\mu$ and $\bar{\nu}_\mu$ production are enhanced, especially for the latter. At such conditions, with electrons being highly energetic,  the formation of muons becomes energetically favorable and negative charge is shared between electrons and muons. Because the muon lepton number is conserved and initially equal to zero, this leads to an increased $\bar{\nu}_\mu$ production. Furthermore, the creation of $\mu^-$ and and $\mu^+$ effectively softens the equation of state of the matter inside the core by converting a fraction of the  thermal and degeneracy energy of electrons into rest-mass energy of muons. This leads to an accelerated shrinking of  the PNS and therefore, to  higher average energies of electron neutrinos, with a consequent increase of neutrino heating. The net result is a higher chance for a successful explosion compared to the case without muon creation \cite{Bollig:2017lki,Bollig:2020phc}.

The goal of our work is to assess the potential impact of muons on the presence of crossings in the $e\mu$, $e\tau$ and $\mu\tau$ sectors. More specifically, we use two sets of data:
\begin{itemize}
\item D1, provided by one-dimensional simulations discussed in Ref. \cite{Bollig:2020} for a 18.6 and 20$M_\odot$ progenitors (adopting the SFHo equation of state \cite{Hempel:2011mk,Steiner:2012rk}), both with and without muons. 
\item D2, presented already in Ref. \cite{Tamborra:2017ubu}, it includes 11, 15 and 25$M_\odot$ progenitors (adopting the LS220 equation of state), where muons were not included.
\end{itemize} 

In these two data sets we look for the presence of ELN crossings through the method proposed in Ref. \cite{Abbar:2020fcl}, which is based on the moments of the neutrino angular distributions. In D1 we find that ELN crossings are relatively common in one-dimensional SN simulations and they occur both below (post-shock region) and above (pre-shock region) the SN shock. This happens regardless of the inclusion of muons, though in their presence crossings are more numerous, or equivalently easier to capture with the moments. Furthermore, the inclusion of muons interestingly allows for a number of ELN crossings in the $\mu\tau$ sector, whereas none are otherwise expected. Adopting the same method in D2 we do not find any crossing neither in the pre-shock nor in the post-shock region (at least at low radial distances). As we explain later in Section V, such a discrepancy with the results of D1 might be due to non equal theoretical inputs in the corresponding numerical models. 

In D2 the neutrino angular distributions are also available, so one can look directly at these for a crossing search. Using this unique opportunity we confirm that crossings occur both above and below the shock. Nevertheless some differences remain with the results obtained with D1.

The structure of the paper is as follows. In Section II we briefly describe the SN models adopted for D1 and D2. In Section III we review the mathematical framework proposed in Ref. \cite{Abbar:2020fcl} to assess the presence of crossings by just using the moments of the neutrino angular distributions. Here we highlight how this method is modified when including muon effects in the supernova model. In Section IV we apply the method previously described to both D1 and D2 and we provide a list of original findings. In Section V we provide our interpretation of the results obtained in D1, D2, from their comparison and we finally draw our conclusions. When appropriate, we also comment on the results obtained in a similar analysis performed in Ref. \cite{Morinaga:2019wsv}.

\section{Supernova Models}

\subsection{D1}
This data contains simulations for two progenitor masses: 18.6 and 20$M_\odot$, already presented in Ref. \cite{Bollig:2020}\footnote{This data is available upon request at \url{https://wwwmpa.mpa-garching.mpg.de/ccsnarchive/data/Bollig2016_angular_moments/}, where the angular moments can be extracted. Radial profiles are available upon request at \url{https://wwwmpa.mpa-garching.mpg.de/ccsnarchive/data/Bollig2016_radial_profiles/}.}. The progenitor models are taken from \cite{Woosley:2002zz} for the 18.6$M_\odot$  case and from \cite{Woosley:2007as} for the 20$M_\odot$ one. The simulations are performed in 1D with the PrometheusVertex code with general-relativistic corrections \cite{Marek:2005if} and six-species neutrino transport, which is solved iteratively through two-moment equations for neutrino energy and momentum, where a Boltzmann closure \cite{Rampp:2002bq} is adopted. The effect of redshift and time dilation are also included in the transport. The tangent-ray method assumes neutrinos to propagate along straight paths instead of curved geodesics, since the effects of ray bending are relatively small. We consider the following set of neutrino reactions (see Appendix A of Ref. \cite{Buras:2005rp} for more details): neutral-current and charged-current interactions with nucleons including nucleon recoils and thermal motions, nucleon-nucleon correlations, weak magnetism, a reduced effective nucleon mass and quenching of the axial-vector coupling at high densities; nucleon-nucleon bremsstrahlung; $\nu\nu$ scattering; and neutrino-antineutrino-pair conversions between different flavors \cite{Buras:2002wt}. In neutrino absorption on nucleons we also take into account the mean-field corrections for the nucleons consistently with the equation of state. Electron capture and inelastic neutrino scattering on nuclei are also included \cite{Langanke_2003,Langanke_2008}. Furthermore, we take into account scattering of all types of neutrinos (i.e. of neutrinos and antineutrinos of all flavors) with muons  and antimuons (see Table I of Ref. \cite{Bollig:2017lki}). A mixing-length approach is employed to simulate PNS convection \cite{Mirizzi:2015eza}. Explosions are artificially triggered for the 18.6$M_\odot$ model, a few 100 ms after bounce, as described in Ref. \cite{Mirizzi:2015eza}. Shock revival occurs and the explosion is artificially triggered in our 1D models when the density of the infalling stellar material above the accretion shock is artificially reduced by a large factor (typically 10 or more). In response to the corresponding sudden decrease of the mass-accretion rate, the shock starts expanding and neutrino heating deposits the energy for an explosion that aborts further accretion and leaves a neutrino-cooling neutron star behind. On the other hand the 20$M_\odot$ model is not exploding. For all progenitor masses simulations are performed with the SFHo \cite{Steiner:2012rk} equation of state. This equation of state is fully compatible with all current constraints from nuclear theory and experiment \cite{Fischer:2013eka,Oertel:2016bki,Fischer:2017zcr} and astrophysics, including pulsar mass measurements \cite{Demorest:2010bx,Antoniadis:2013pzd,Cromartie:2019kug} and the radius constraints deduced from gravitational-wave and Neutron Star Interior Composition Explorer measurements \cite{Abbott:2018exr,Bauswein:2017vtn,Essick:2020flb}. To provide a better comparison with D2 (see next subsection) the simulation of the 18.6$M_\odot$ model is performed also with the LS220  \cite{Lattimer_LS220} equation of state.

\subsection{D2}
This data contains models for three progenitor masses of $11.2$, $15$ and $25M_\odot$\footnote{This data is publicly available at \url{https://wwwmpa.mpa-garching.mpg.de/ccsnarchive/data/Tamborra2017/}}. The progenitor models are taken from \cite{Woosley:1995ip} for the $15M_\odot$ case and from \cite{Woosley:2002zz} for all other cases. The nuclear equation of state is LS220 from \cite{Lattimer_LS220} with compressibility modulus $K=220$ MeV \cite{Hudepohl:2013zsj}. The simulations were performed with the 1D version of the same (Prometheus-Vertex) code employed for D1. Therefore the method for solving the hydrodynamical evolution and neutrino transport is the same as for model set D1, as well as the set of neutrino interactions accounted for\footnote{In the charged-current neutrino-nucleon interactions, the simulations in D2 did not yet consider the nucleon mean-field potentials.}. The main differences with D1 are the following: here only three-neutrino species are considered ($\nu_e,\bar{\nu}_e$ and $\nu_x$) and neither muon physics nor proto-NS convection are included. Moreover, no artificial explosion is triggered. 

\section{Analysis method}

Under the assumption of $\nu_x = \bar\nu_x$ and azimuthal symmetry in the neutrino gas,and working with natural units $\hbar=c=1$, an ELN crossing occurs when
\begin{equation}
G_{e}(\gamma)=\sqrt{2}G_F\int_0^\infty\frac{dEE^2}{2\pi^2}[f_{\nu_e}(E,\gamma)-f_{\bar{\nu}_e}(E,\gamma)]\,,
\label{eq:fast_conversions_condition}
\end{equation}
changes its sign at a given $\gamma_0$. 
Here, $\gamma=\cos\theta$ with $\theta$ being the neutrino propagation angle with respect to the radial direction and $f_{\nu_\alpha}$ is the neutrino occupation number for a given flavor $\alpha$. The quantity $G_{e}$ physically represents the angular distribution of the ELN. However, the inclusion of muon physics inevitably introduces a non-negligible difference between $\nu_\mu$ and $\bar{\nu}_\mu$. In this case, the relevant quantities to be taken into account are \cite{Capozzi:2020kge}
\begin{eqnarray}
G_{e\mu}(\gamma)&=&G_{e}(\gamma)-G_{\mu}(\gamma)\,,
\label{eq:fast_conversions_condition_muons_1}\\
G_{e\tau}(\gamma)&=&G_{e}(\gamma)-G_{\tau}(\gamma)\,,
\label{eq:fast_conversions_condition_muons_2}\\
G_{\mu\tau}(\gamma)&=&G_{\mu}(\gamma)-G_{\tau}(\gamma)\,,
\label{eq:fast_conversions_condition_muons_3}
\end{eqnarray}
where $G_\mu$ and $G_\tau$ are defined equivalently to $G_e$ in Eq.~(\ref{eq:fast_conversions_condition}), but for $\nu_\mu$ and $\nu_\tau$, respectively. Analogously to $G_{e}$, $G_{\alpha\beta}$ represents the angular distribution of the difference between the lepton numbers for flavor $\alpha$ and $\beta$. Considering that in general $f_{\nu_\mu} < f_{\bar{\nu}_\mu}$, some differences with the standard calculations in the absence of muons should be expected. Nevertheless, whether the quantities in Eqs.~(\ref{eq:fast_conversions_condition_muons_1})-(\ref{eq:fast_conversions_condition_muons_3}) can change sign has never been studied so far, mainly because the differences between $\nu_\mu$ and $\bar{\nu}_\mu$ are usually thought to be not comparable to those between $\nu_e$ and $\bar{\nu}_e$. Moreover, until very recently no CCSN simulations has included muons in the equation of state and neutrino transport. The first  two-dimensional simulations in this direction have been presented in Ref.~\cite{Bollig:2017lki}.

Usually, multidimensional simulations of CCSNe do not provide full angular information of  $G_{\alpha\beta}(\gamma)$, since the equations for neutrino transport are solved for a number of the moments of  $G_{\alpha\beta}(\gamma)$, which are defined as
 \begin{equation}
I^{\alpha\beta}_n = \int_{-1}^{1} \mathrm{d}\gamma\ \gamma^n\ G_{\alpha\beta}(\gamma).
\label{eq:moments}
\end{equation}
In Ref. \cite{Abbar:2020fcl}, a simple method was proposed to assess the presence of crossings in the angular distribution of $G_{\alpha\beta}(\theta)$. This method attempts to find a positive function $\mathcal{F}(\gamma)>0$, such that the quantity
\begin{equation}
I_\mathcal{F}^{\alpha\beta} = \int_{-1}^{1} \mathrm{d}\gamma\ \mathcal{F}(\gamma)\  G_{\alpha\beta}(\gamma)
\label{eq:IF}
\end{equation}
satisfies the following relation
\begin{equation}
I^{\alpha\beta}_\mathcal{F} I_0^{\alpha\beta} < 0\,.
\label{eq:crossing_condition}
\end{equation}
 If such a function exists, then $G_{\alpha\beta}(\gamma)$ is assumed to have a crossing. In order to write $I^{\alpha\beta}_\mathcal{F}$ in terms of the available moments $I^{\alpha\beta}_n$, defined in Eq.~(\ref{eq:moments}), we choose $\mathcal{F}(\gamma)$ to be a polynomial of $\gamma$
 \begin{equation}
 \label{eq:F}
\mathcal{F}(\gamma) =\sum_{n=0}^N a_n \gamma^n,
\end{equation}
where $N$ is the highest  moment available from numerical simulations and $a_i$'s are an arbitrary set of coefficients for which $\mathcal{F}(\gamma)>0$. According to this choice, $I^{\alpha\beta}_\mathcal{F}$ becomes
\begin{equation}
I^{\alpha\beta}_\mathcal{F}=a_0I^{\alpha\beta}_0+a_1I^{\alpha\beta}_1+a_2I^{\alpha\beta}_2+\dots+ a_NI^{\alpha\beta}_N\,.
\label{eq:I_F}
\end{equation}
Note that in supernova simulations the (infinite) system of moment equations of the Boltzmann transport equation (obtained as angular integrals of different orders) is often cut at the level of the neutrino momentum (first-order moment) equation through the assumed algebraic closure relations.\footnote{We point out that this does {\em not} apply to the VERTEX code used in computing the supernova models analysed in the present paper.  VERTEX solves the first and second-order moment equations iteratively for convergence with a model Boltzmann equation. It thus obtains angular distribution functions as well as representations of the angular moments of second order and higher as solutions of the Boltzmann equation. Therefore VERTEX does not employ any ad hoc assumptions for the closure relations.} In general we have $N\leq3$. The results reported herein refer to $N=2$, i.e. using quadratic polynomials for $\mathcal{F}(\gamma)$. We believe the results with $N=3$ are less reliable, since the third order moment is the most severely affected by the ad hoc prescriptions of closure relations. In principle one can even use only linear polynomials. However, in this case a significant amount of information is lost and, consequently, not many crossings can be found. For the sake of completeness, we show in the Appendix a direct comparison between the performance of our method with different orders for the polynomials.

\section{Analysis Results}

\begin{figure*}
  \centering
    \includegraphics[width=0.45\textwidth]{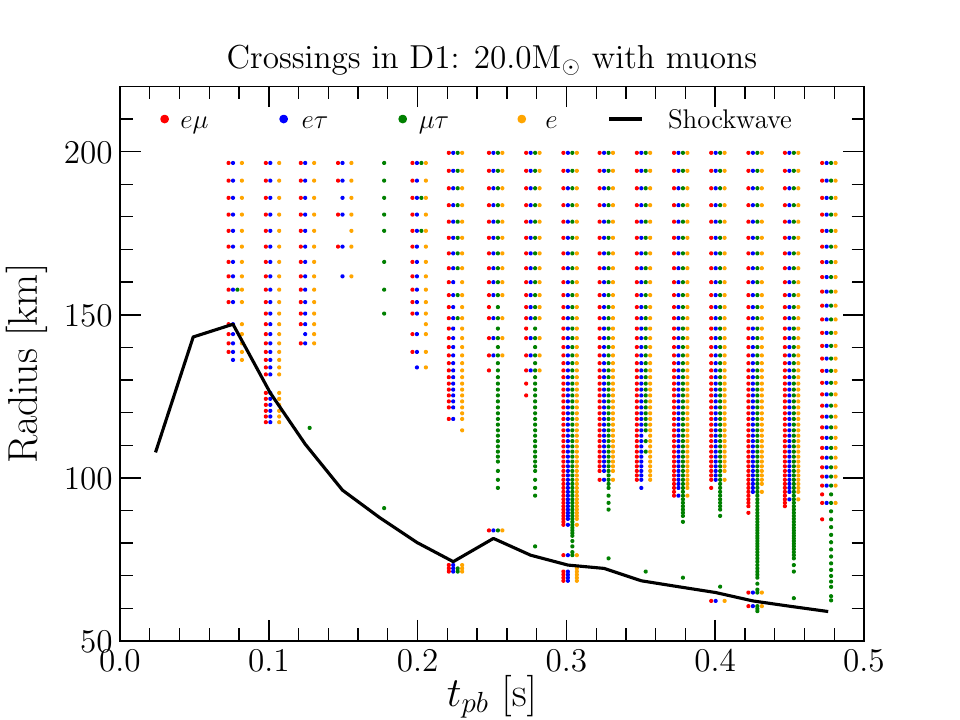}
          \includegraphics[width=0.45\textwidth]{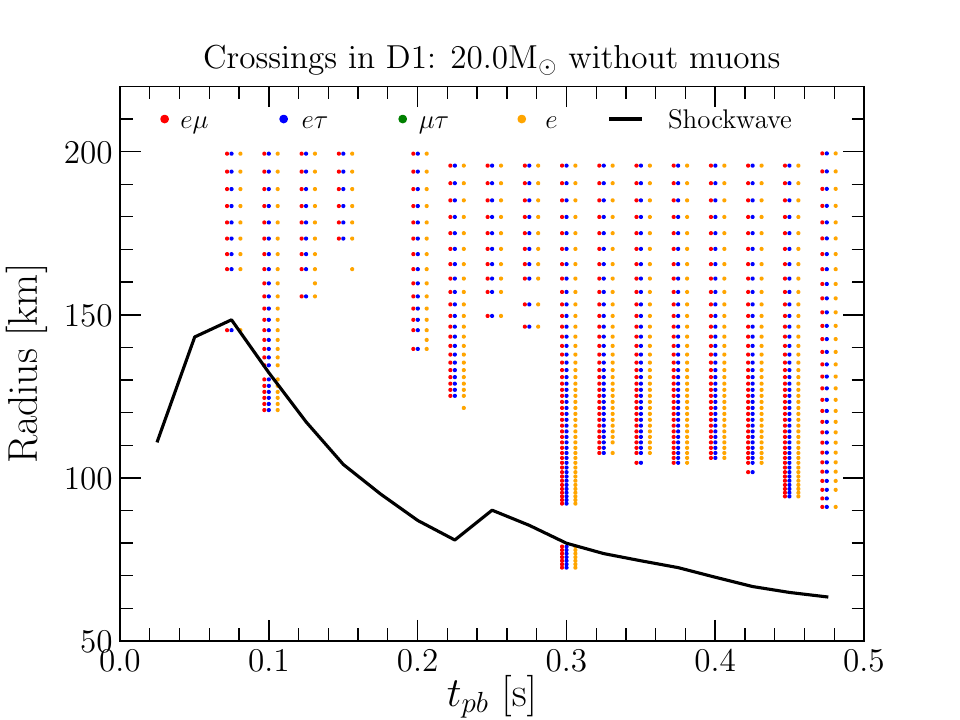}
 \includegraphics[width=0.45\textwidth]{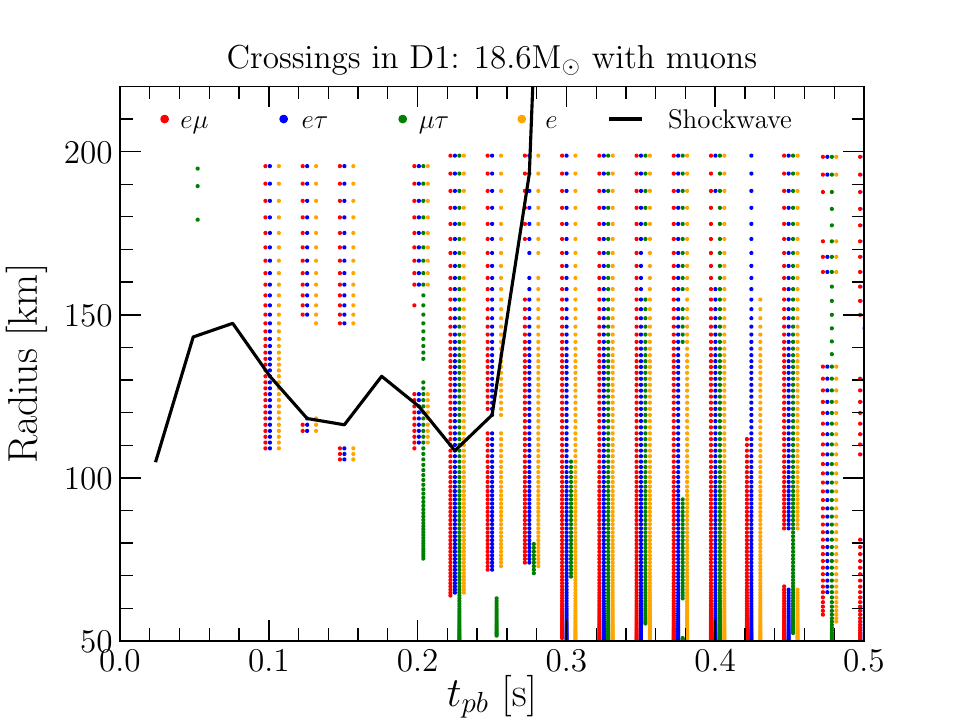}
 \includegraphics[width=0.45\textwidth]{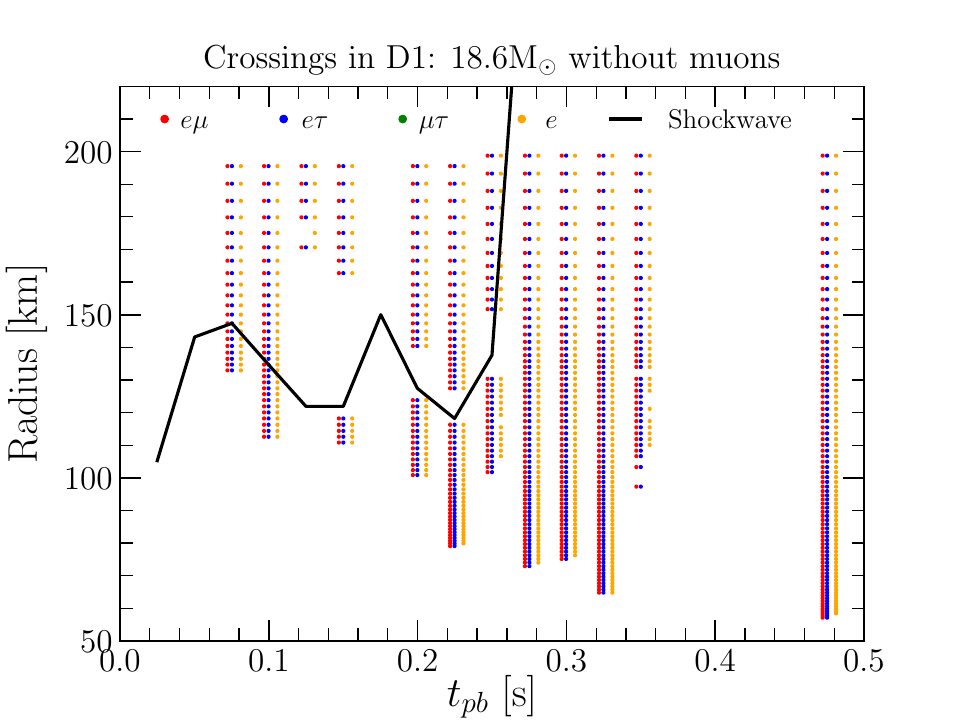}
       \caption{Crossing points found in D1 with the method proposed in Ref. \cite{Abbar:2020fcl}, as a function of time and radial distance. The red, blue, green and orange dots refer to crossings for $G_{e\mu},G_{e\tau},G_{\mu\tau}$ and $G_e$, respectively. In order to  have a clearer separation between different dots, they are slightly shifted horizontally from their correspondent post bounce time. The black line indicates the evolution of the SN shock in time and space.}
        \label{fig:summary_plot_D1}
\end{figure*}

Apart from observing crossings in the $\mu-\tau$ sectors, we find a number of crossings below the SN shock in the region close to the shock front. This is indicated in Fig.~\ref{fig:summary_plot_D1}, where with colored dots the post bounce times (x-axis) and radial distances from the center of the SN (y-axis) are shown for which we find crossings in D1,  assuming $\mathcal{F}(\gamma)$ to be a quadratic polynomial. The top (bottom) row refers to the 20 (18.6)$M_\odot$ model. The left (right) column refers to the case with (without) muons. The red, blue, green and orange dots refer to crossings for $G_{e\mu},G_{e\tau}$, $G_{\mu\tau}$ and $G_{e}$, respectively. Note that, in order to  have a clearer separation between different dots, the red, green and orange ones are slightly shifted horizontally from their correspondent post bounce time. The black line indicates the evolution of the shock in time and space. Obviously, the 20$M_\odot$ model is not exploding, whereas the 18.6$M_\odot$ case is exploding, where the explosion was triggered  artificially  around $t_{pb} \approx 0.26$~s. All times in our model analysis are given as post-bounce times, $t_{pb}$, i.e. as times after core bounce, because the core bounce effectively coincides with the time when the supernova shock is launched. The start of the explosion (when present) is, of course, not the time when the shock wave is produced, but the time when the stalled shock wave is revived.

A few remarks are in order. The radial distance on the $y$-axis is limited to $[50,200]$ km because below this range no crossings are found, whereas above it they are ubiquitous. Moreover, since the spatial grid changes as a function of time, the upper dots in each time snapshot might not be horizontally aligned, but this does not mean that crossings do not occur above 200 km for some snapshots.

\begin{figure*}
  \centering
  \includegraphics[width=0.45\textwidth]{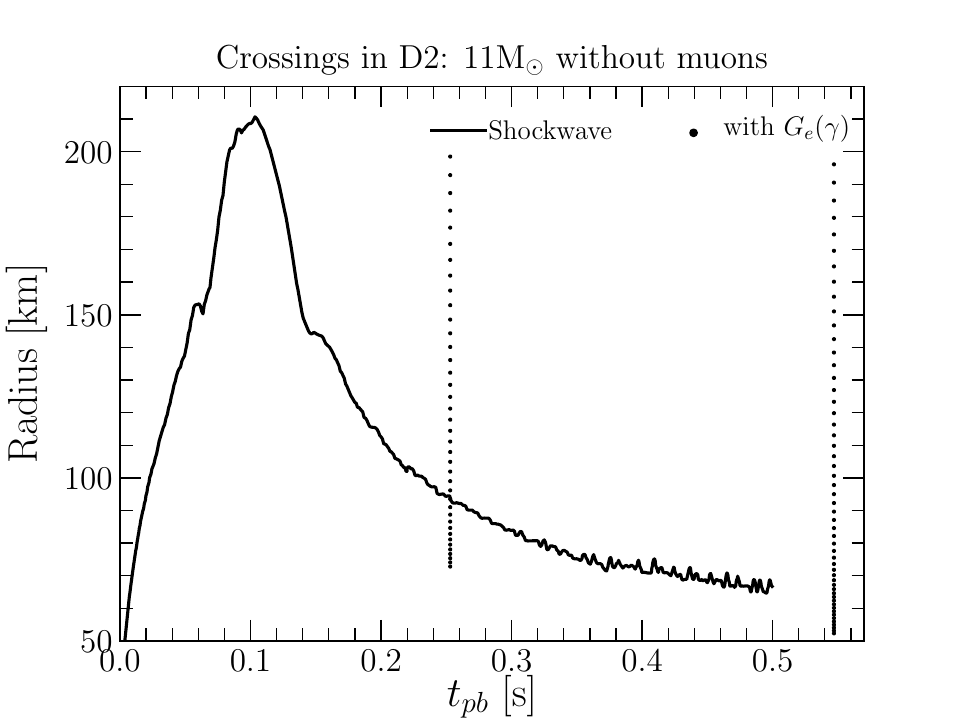}
    \includegraphics[width=0.45\textwidth]{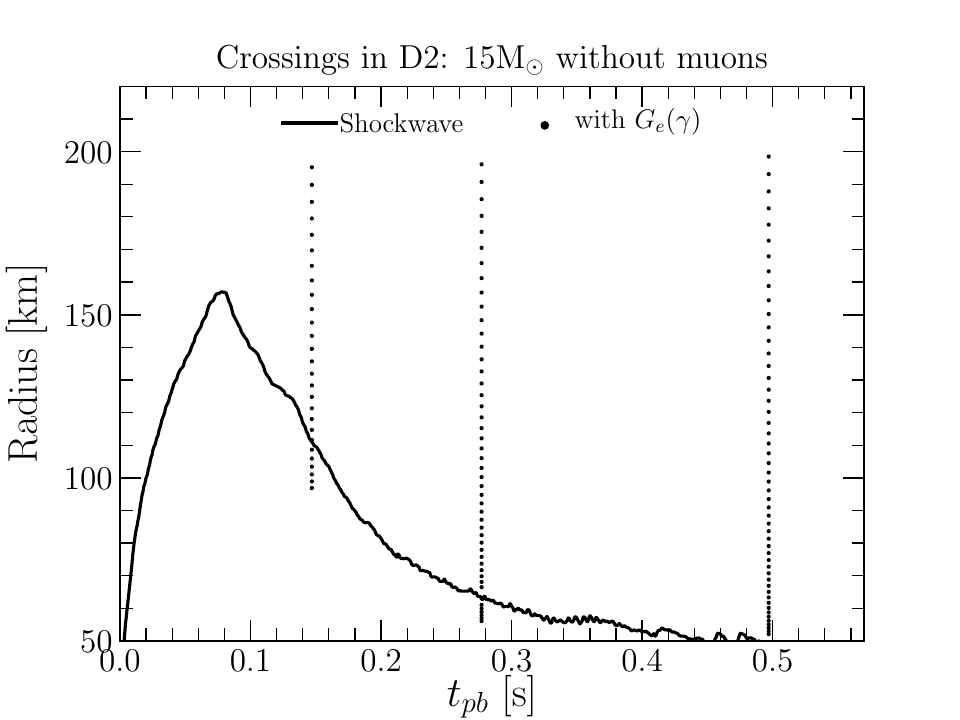}
    \includegraphics[width=0.45\textwidth]{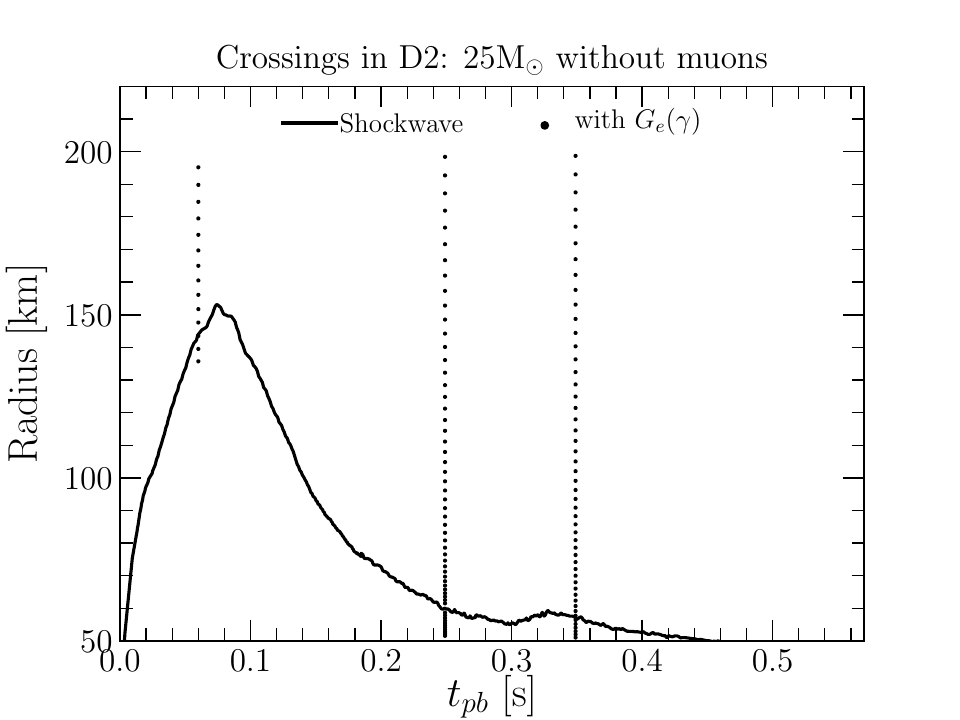}
       \caption{Same as Fig. \ref{fig:summary_plot_D1} but for the crossings in the 15$M_\odot$ of D2.
       Note that the crossings can be captured here by looking directly at the ELN angular distribution 
       (labeled as ``with $G_e(\gamma)$"). With the moments method crossings are found only with cubic polynomials for the radial distance range considered in the plots. These are not reported here, since we are only showing results connected to quadratic polynomials.}
        \label{fig:summary_plot_D2}
\end{figure*}

Fig.~\ref{fig:summary_plot_D2} is equivalent to Fig.~\ref{fig:summary_plot_D1}, but it refers to the (not exploding) 11, 15 and 25$M_\odot$ models of D2, which do not include muons. Note that only three time snapshots are available in the public version of the data. The crossings  shown in Fig. \ref{fig:summary_plot_D2} are observed only by directly looking at the angular distributions, such as those displayed in Fig. \ref{fig:angular_distributions}. No crossings are found with quadratic polynomials within the radial distance range used in the plots. As shown in the Appendix (see Fig. \ref{fig:summary_plot_D2_with_cubic_polynomials}), only cubic polynomials allow crossings to be found at $r<200$ km, but none below the shock.

\begin{figure*}
  \centering    
    \includegraphics[width=0.98\textwidth]{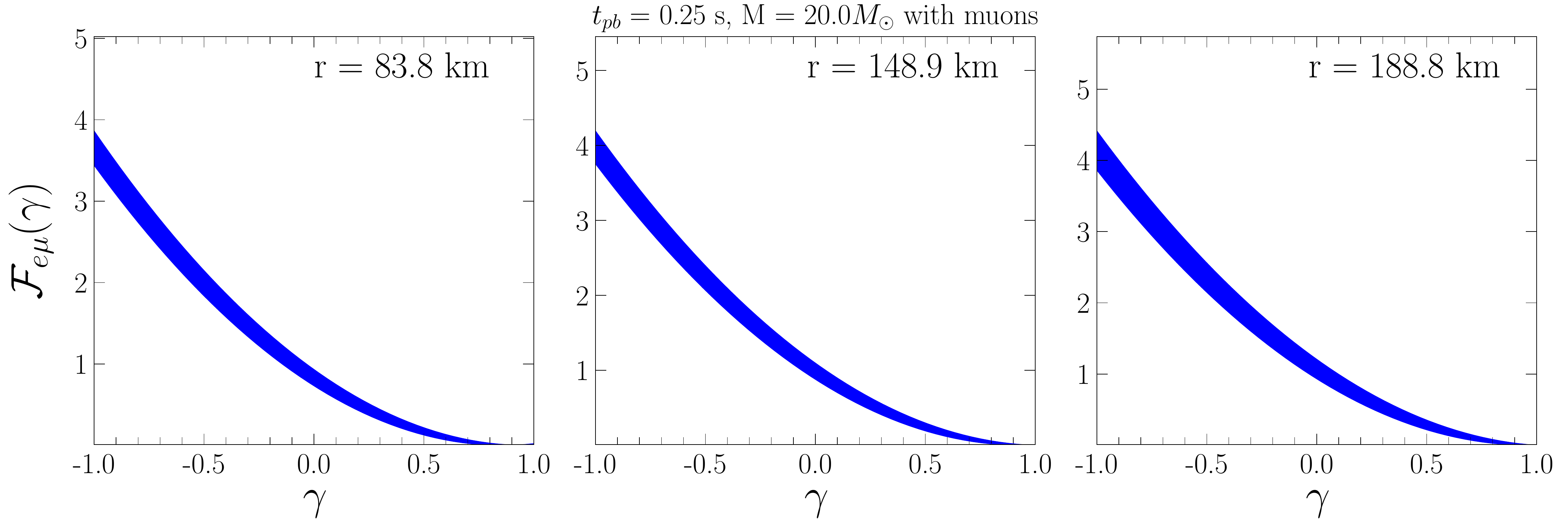}
     \includegraphics[width=0.98\textwidth]{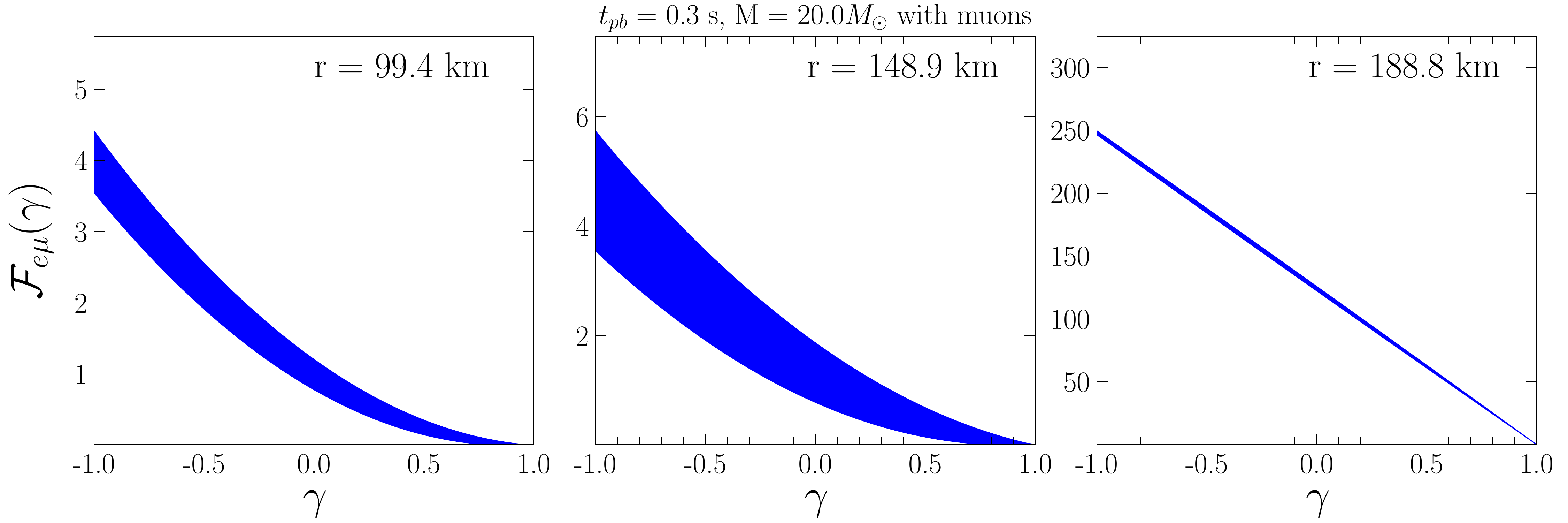}
       \caption{Shape of \ $\mathcal{F}_{e\mu}(\gamma)$ for which crossings are found at $t_{pb}=0.25$ s (top panels) and $t_{pb}=0.3$ s (bottom panels) for three different radial distances. Here we have considered the 20$M_\odot$ model with muons of D1. $\mathcal{F}(\gamma)$ is assumed to be a quadratic polynomial.}
        \label{fig:F_shape}
\end{figure*}

\begin{figure*}
  \centering
    \includegraphics[width=\textwidth]{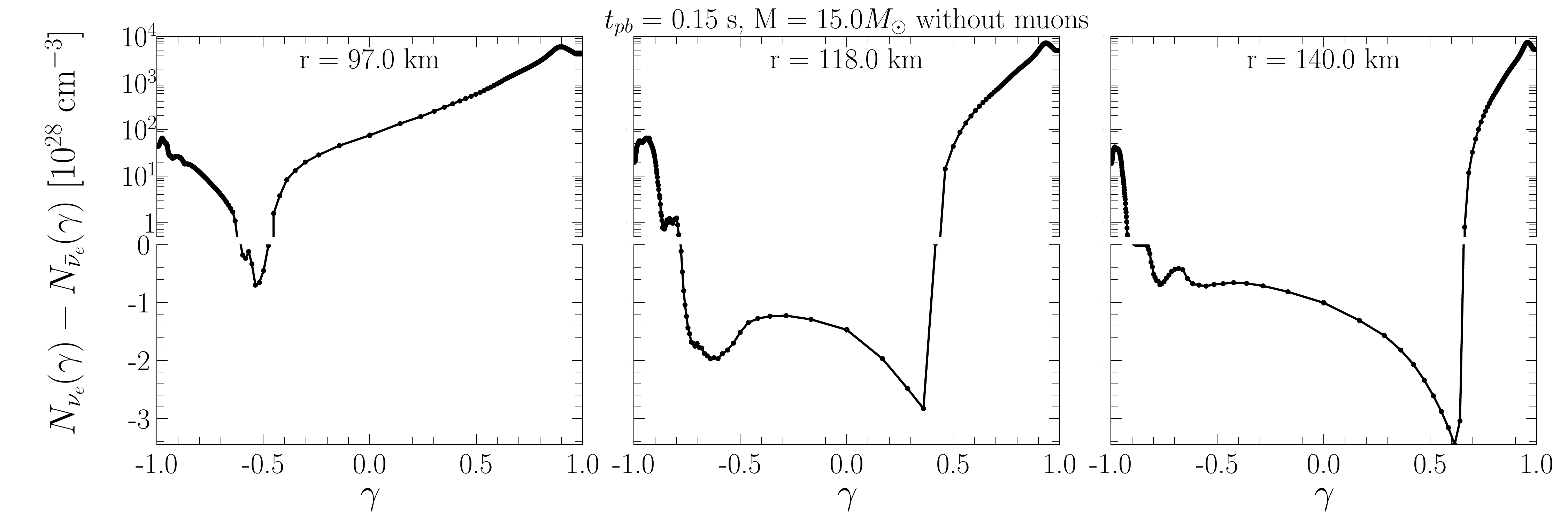}
       \caption{Angular distributions of ELN
       taken from the publicly available D2. In particular, these three panels represent three radial distances where crossings are found in the 15$M_\odot$ model at 0.15 s.}
        \label{fig:angular_distributions}
\end{figure*}

Fig.~\ref{fig:F_shape} displays the shape of $\mathcal{F}(\gamma)$ that allows to capture the crossings in D1. Since for each crossing there might be multiple allowed combinations of the polynomial coefficients, the shape is presented by filling the area between the lowest and highest values of $\mathcal{F}$.  In all cases the functional form is assumed to be a quadratic polynomial. For the sake of brevity, this figure only contains information regarding the 20$M_\odot$ model with muons of D1 at $t_{pb}=0.25$ s (top panels) and $t_{pb}=0.30$ s (bottom panels). For each time snapshot we  plot the shape of $\mathcal{F}(\gamma)$ for three radial distances. The same shape of $F(\gamma)$ is obtained in the $e\tau$ and $\mu\tau$ sectors and for all other time snapshots, radial distances and models. More specifically we find that the function is always vanishing at $\gamma\to1$ and it reaches its maximum at $\gamma\to-1$.
 
 Fig.~\ref{fig:angular_distributions} displays the shape of the angular distribution uniquely provided by D2. We have selected only three radial distances where crossings are present in the 15$M_\odot$ model at 0.15 s.  
 
The most important original findings stemming from Figs. \ref{fig:summary_plot_D1}-\ref{fig:angular_distributions} are:
\begin{enumerate}
\item Starting from $t_{pb}\simeq 0.08$~s, there are crossings occurring below the SN shock in D1. 
\item At fixed post bounce time, above the SN shock location there is usually a gap in terms of radial distance, where no crossing is found. The width of such a gap is time dependent and it nearly disappears when the explosion sets in the 18.6$M_\odot$ model of D1. The gap is absent when finding crossings using directly the angular distributions. 

\item There exist post bounce times and radial distances for which only crossings in the $\mu-\tau$ sector are present. This is the first time this possibility is demonstrated in a supernova simulation. By construction, such crossings cannot exist without considering muons, as usually done in the literature, since in this case $G_\mu(\gamma)=G_\tau(\gamma)$.
\item There is a hint that all the captured crossings are in the backward direction. This stems from the fact that $\mathcal{F}(\gamma)$ is maximal at $\gamma=-1$ and vanishing at $\gamma=+1$. Such a result is in agreement with what was found in \cite{Morinaga:2019wsv}. The maximal value stays nearly constant when changing the radial distance. However, for some time snapshots we observe a sudden increase by a few orders of magnitude.
\item A larger number of crossings are present in the $e\mu$ and $e\tau$ sectors in the presence of muons (left panels of Fig. \ref{fig:summary_plot_D1}). Crossings below the SN shock do not disappear in the absence of muons in the same model.
\item We find crossings in D2, whereas none were reported in Ref. \cite{Tamborra:2017ubu}, where the same data has been analyzed. Indeed  the focus of that work was to look for deep crossings in the forward directions. However, such crossings are found only by looking directly at the angular distributions and not through the moments method, unless cubic polynomials are employed. This is in contrast with D1, where quadratic polynomials are enough.
\item The angular distributions in Fig. \ref{fig:angular_distributions} suggest that, if present, crossings come at least in pairs and that the region in $\gamma$ where $G_{e}$ becomes negative gets larger with radial distance. 

\end{enumerate}

A more detailed explanation is in order concerning why we believe the crossings are in the backward direction. This consideration stems from Fig.~\ref{fig:F_shape}, since the shape of $\mathcal{F}(\gamma)$ may provide some hints concerning the shape of $G_{\alpha\beta}(\gamma)$, even if one has no direct access to it. In our case $\mathcal{F}(\gamma)$ is monotonically decreasing from its maximum value at $\gamma=-1$ to 0 at $\gamma=+1$. In order to satisfy Eq.~\ref{eq:crossing_condition}, $\mathcal{F}(\gamma)$ must suppress $G_{\alpha\beta}$ for those values of $\gamma$ where it is the largest, i.e. where the largest contribution to the moment integral $I_0^{\alpha\beta}$ (and to its sign) is coming from. Conversely,  $\mathcal{F}(\gamma)$ must enhance $G_{\alpha\beta}$ where it assumes the smallest values (and the opposite sign). Indeed this is the only way to make $I_{\mathcal{F}}^{\alpha\beta}$ and $I_0^{\alpha\beta}$ have opposite signs, thus satisfying the crossing condition in  Eq. \ref{eq:crossing_condition}. Consequently, given the shape of $\mathcal{F}$ we expect $G_{\alpha\beta}$ to have a specific sign for a narrow region close $\gamma=+1$ and the opposite one for a large range of $\gamma$ extending from $\gamma=-1$ to (probably) some $\gamma>0$. This mean that $G_{\alpha\beta}$ changes sign when passing from the forward to the backward direction. There are two possibilities: $G_{\alpha\beta}$ is large and positive in the forward direction and small and negative in the backward direction, or vice versa. Such degeneracy can be solved by considering that supernova neutrinos have a larger flux compared to the antineutrinos and that all the crossings are in the free-streaming regime, where the angular distributions are forward peaked.

\section{Interpretation of the results}

In this section our goal is to provide tentative explanations for the results listed above. We emphasize that, given a set of moments of the neutrino angular distributions, our analysis method can only establish whether a crossing is present or not at a given point in space and time. On the other hand, the underlying physics of crossings can only be tentatively inferred on grounds of plausibility. For instance, this can be done by taking into account the shape of $\mathcal{F}(\gamma)$ or well-understood facts and conditions of the physics of supernovae\footnote{Since the VERTEX code provides angular neutrino distributions locally, an in-depth analysis of the neutrino phase space in energy and angle would be possible in principle. However, tracking down the physics reasons of detected crossings would require to also inspect the energy-dependent effects of different neutrino interactions in different stellar regions. The corresponding post-processing of the simulation results requires a huge effort and is beyond the scope of the present work, which is supposed to diagnose the fundamental differences between supernova conditions with and without the inclusion of muon physics.}. Therefore, the explanations we propose must be considered only as suggestions, since we do not have any final confirmation from a rigorous analysis, which is beyond the scope of our work.

Until now, no crossings have ever been reported in the post-shock region of one-dimensional CCSN simulations~\cite{Tamborra:2017ubu,Morinaga:2019wsv}, though muons have never been taken into account. In particular in Ref.~\cite{Tamborra:2017ubu} no crossings were reported, whereas in Ref.~\cite{Morinaga:2019wsv} they were found only in the SN pre-shock region, and associated with an enhanced coherent scattering of the more energetic $\bar{\nu}_e$ on heavy nuclei. The difference in scattering probability generates a backward crossing. Our results, both with and without muons, confirm the findings of Ref.~\cite{Morinaga:2019wsv}. Indeed, in Figs.~\ref{fig:summary_plot_D1} and \ref{fig:summary_plot_D2}, most of the crossings are in the pre-shock region. Moreover we know from Fig.~\ref{fig:F_shape}  that $\mathcal{F}(\gamma)$ has a peak at $\gamma=-1$, a feature most likely related to a backward crossing.

There is one characteristic of the shape of $\mathcal{F}(\gamma)$ in Fig.~\ref{fig:F_shape} we are not able to fully explain. For $t_{pb}=0.3$ s and a few other time snapshots (not shown), there is a sudden and significant enhancement of the maximal value of $\mathcal{F}(\gamma)$ when increasing the radial distance. Conversely, for most time snapshots the maximal value seems to be slowly changing with the radial distance. The sudden enhancement suggests that the features of $G_{\alpha\beta}(\gamma)$ might change significantly when moving away from the core of the supernova. Therefore, the mechanism generating the crossing might be modified at specific locations of space and time, though the backward nature (and possibly the connection with some sort of scattering) stays unaffected. We are unable to advance any hypothesis about what kind of mechanism change is involved.

Our reanalysis of D2~\cite{Tamborra:2017ubu} supports the presence of backward crossings both below and above the shock. As Fig.~\ref{fig:angular_distributions} shows, the sign of the ELN changes from positive to negative when going from $\gamma=+1$ towards $-1$. However, an extra crossing is present at lower values of $\gamma$. We are unable to assess a possible origin of such a feature.

If one applies the shape of $\mathcal{F}(\gamma)$ displayed in Fig.~\ref{fig:F_shape} to a $G_{\alpha\beta}$ with the double crossing feature we just described, most likely Eq.~\ref{eq:crossing_condition} would not be satisfied. Indeed, $\mathcal{F}(\gamma)$ needs to be nearly vanishing at both $\gamma=\pm1$, whereas in Fig.~\ref{fig:F_shape} it happens only at $\gamma=+1$. Thus, we do not expect angular distributions like those in Fig.~\ref{fig:angular_distributions} to be occurring in D1. This might be related to different sets of theoretical ingredients used in the SN simulations, as we discuss in more details later in this Section.

Another interesting observation is the presence of crossings in the $\mu-\tau$ sector, which can be explained as follows. Both the number densities and the average energies of $\nu_\mu$ are smaller than those of $\bar{\nu}_\mu$, but the neutral-current cross section for the former is larger assuming the same energy. This is due to weak magnetism corrections~\cite{Horowitz:2001xf} and this is enough to compensate the not so large energy difference between $\nu_\mu$ and $\bar{\nu}_\mu$, which is usually not larger than ~0.5 MeV, with $\bar{\nu}_\mu$ being more energetic. This means that $\bar{\nu}_\mu$ should be dominating over $\nu_\mu$ in the forward direction, whereas the opposite is happening in the backward one. In principle,  the $\tau$ flavor can also play a role. But  given the fact that the  number densities of $\nu_\tau$ and $\bar{\nu}_\tau$ are very similar,  the contribution from $G_{\tau}(\gamma)$ should be negligible in $G_{\mu\tau}$.

\begin{figure*}
  \centering
   \includegraphics[width=0.45\textwidth]{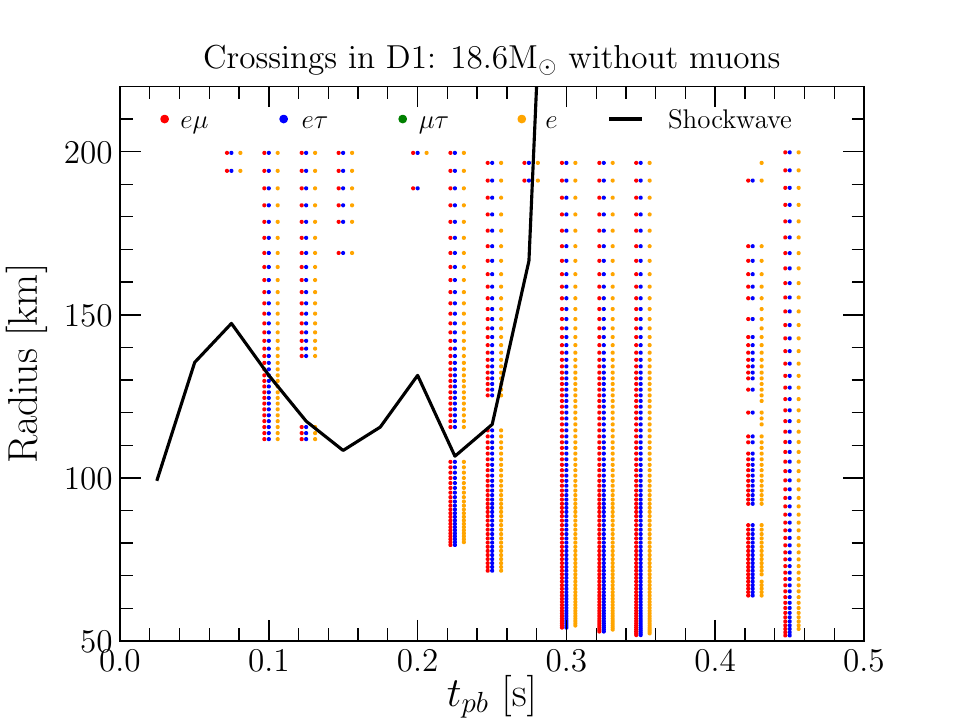}
       \caption{Same as the lower right panel of Fig.~\ref{fig:summary_plot_D1}, but for the LS220 equation of state (which was also used in the D2 set of models) instead of the SFHo.}
        \label{fig:summary_plot_D1_LS220}
\end{figure*}

An intriguing aspect of our results is that some crossings are found with the moments method in the post-shock region of D1. None of these were found in Ref.~\cite{Morinaga:2019wsv}, whereas in D2 they can be highlighted only by looking directly at the angular distributions and not by employing the moments. This suggests that post-shock crossings may indeed be present in all models, but they are more pronounced and easier to find  in D1. This aspect is not related to the absence of muons in the previous works. Indeed, in D1 we find crossings below the SN shock both in the pre- and post-shock regions, regardless of the inclusion of muons. The reason for this difference compared to the previous studies might be related to non-equal set of theoretical inputs in the numerical models. A first possibility is the different equation of state of the PNS. Indeed, this has an influence on neutrino luminosities, neutrinosphere temperatures, mean neutrino energies and the PNS radius. For instance D1 and D2 are obtained with the following equation of states: SFHo for the former and LS220 \cite{Lattimer_LS220} for the latter. However, we have checked the data from a version of D1 with LS220, with all the other inputs unchanged,  and found that the pattern of crossings remains basically the same. This is shown in Fig.~\ref{fig:summary_plot_D1_LS220}. A second possibility might be connected  to PNS convection, which is included in all D1 models through a mixing-length approximation, but not considered in D2. Convection has the following consequences: it increases the PNS radius, it increases the $\nu_e,\nu_\mu$ and $\nu_\tau$ luminosities but reduces the electron $\bar{\nu}_e$ luminosity relative to $\nu_e$ \cite{Buras:2005tb,Nagakura:2019tmy}. It also changes the $\nu_e$ and $\bar{\nu}_e$ spectra relative to each other. Another possibility consists in different inputs in terms of microphysics. Indeed, the models in D1 and D2 use different neutrino reactions than those employed in Ref. \cite{Morinaga:2019wsv}. The hypotheses listed before might represent also the reason why D2 has double crossings and, consequently, a different shape of $\mathcal{F}(\gamma)$ compared to the one seen in D1.

The last aspect worth mentioning is the presence of the gap just above the SN shock, i.e. a range of radial distances for which no crossings is found with the moments method. Most likely here crossings are present, but they are harder to find since inaccessible higher order moments are required in order to capture them. Indeed, if we look directly at angular distributions in D2 we find a continuous set of crossings starting from a radial distance in the post-shock region: the smaller the distance the narrower the crossing. Another proof is that if we pass from cubic to quadratic polynomials for $I_\mathcal{F}^{\alpha,\beta}$ most of the crossings disappearing are those at the low radial distances above the SN shock, as can be seen comparing the middle and right panel of Fig. \ref{fig:polynomials_comparison} in the Appendix. Possible explanations of why crossings are harder to find in the gaps are the following. First, in the pre-shock region the fraction of heavy nuclei is increasing with the distance and consequently also the size of the crossings in the backward directions does. This possible connection is physically related to the fact that the temperatures in the pre-shock flow at small radii become large enough to dissociate iron-group nuclei to alpha particles, thus reducing coherent scattering cross sections. We have directly verified in the data that the fraction of heavy nuclei is slowly increasing after the shock wave, though the rapidity of variation depends on the time snapshot. Moreover, the gap width is smaller in the 18.6$M_\odot$ model of D1 especially after the explosion has set in. Under these circumstances, since the SN shock propagate outwards, the temperature in the pre-shock region is not high enough to produce a significant dissociation of heavy nuclei. 

\section{Conclusions}
A full understanding of the outcome of neutrino flavor conversions in CCSN is currently missing. Such an achievement would represent a milestone in both particle physics and astrophysics, since the next SN neutrino burst can be correctly interpreted only when the flavor composition at the source is precisely reconstructed. A step towards this ambitious goal consists in assessing whether the condition for fast conversions, i.e. a crossing in the angular distribution of the neutrino lepton number, is satisfied in realistic SN models. In this paper we have considered state-of-the-art one-dimensional CCSN simulations where muons is taken into account. In this context, a full three-flavor approach is mandatory, since $\nu_\mu$ and $\nu_\tau$ behave in different ways. We have concluded that muons has only a mild impact on the generation of crossings, though
it seems to make the crossings more pronounced.

We confirm that backward crossings are relatively common above the SN shock, as first pointed out in Ref.~\cite{Morinaga:2019wsv} and recently confirmed in the context of 3D simulations \cite{Abbar_3D}. However, we emphasize that they are also most likely  common in the post-shock region, too, but they are not always observed  due to their dependence on  different inputs used in CCSN simulations. In particular, PNS convection is taken into account  in D1~\cite{Mirizzi:2015eza} but not in D2 \cite{Tamborra:2017ubu}. The consequences of PNS convection on the neutrino emission properties might enhance the depth of crossings, thus making them less difficult to be  observed with  the moments method. Such a hypothesis needs to be verified by a direct comparison of two simulations with and without convection regarding the otherwise exactly same model. However, some confidence on the validity of our conclusion comes from the results of \cite{Abbar_3D}, which show crossings below the shock in 3D simulations. But also in this case, confirmation from other 3D models is required for a conclusive assessment.
 
Another original outcome of our analysis is the observation of  crossings  in the $\mu\tau$ sector, whereas none have been reported so far in other simulations because of their equivalent treatment of neutrinos of the muon and tau flavors. Intriguingly, there are locations in time and space where such crossings are found, where none are observed in the $e\mu$ and $e\tau$ sector. In analogy to all other crossings, these are also in the backward direction and are related to those neutrino interactions that can still occur in the free-streaming regime.

We underline once more that the interpretation of our results must be treated as only suggestive of the physics connected to the observed crossings. Nevertheless, our findings support the need for performing a similar analysis in the context of other one-dimensional models, with different sets of theoretical ingredients. Such a systematic study can augment the understanding we tried to develop in this paper. In the end, this will lead to an improved understanding of the creation of angular crossings. Moreover, our results require a conclusive confirmation from multi-dimensional models with muons. Two-dimensional ones have been already presented in Ref. \cite{Bollig:2017lki} and the results from first three-dimensional simulations including muons have recently been published in Ref. \cite{Bollig:2020phc}.

Even if one assumes that the results obtained in the present work hold in the multi-dimensional models, too, the question of what is the impact of backward crossings remains unanswered. The backward crossings found in this study seem to exist only slightly below the SN shock.  Nevertheless, more crossings may be present even deeper in the star, though they would be significantly more shallow. Although the occurrence of this backward ELN crossings can lead to fast instabilities with significant growth rates, it is not yet clear whether they can lead to significant flavor conversions. Indeed, these instabilities mostly affect (in the linear regime) the backward traveling neutrinos (see Fig.~7 in Ref.~\cite{Abbar_3D}), which are only a very tiny fraction of the neutrino population. The ultimate assessment must come through numerical simulations. One can perform simplified simulations of neutrino flavor evolution, completely decoupled from the hydrodynamics of the SN. On the other hand, a more reliable answer must come from fully including flavor conversions in self-consistent  simulations of CCSNe, or treating them with an effective approach in the same context, which is computationally more affordable. We hope our work will trigger future endeavors in this direction.

\section*{Acknowledgments}
We thank Irene Tamborra for providing the numerical tables for the shockwave evolution in the three models of D2. 

F.C., S.A., R.B. and H.T.J. acknowledge partial support by the Deutsche Forschungsgemeinschaft (DFG, German Research Foundation) through Grant Sonderforschungsbereich (Collaborative Research Center) SFB-1258 ``Neutrinos and Dark Matter in Astro- and Particle Physics (NDM)''. FC's work at Virginia Tech is supported by the U.S. Department of Energy under the award number DE-SC0020250 and DE-SC0020262. Additionally, at Garching, funding by the European Research Council through Grant ERC-AdG No.~341157-COCO2CASA and by the DFG under Germany's Excellence Strategy through Cluster of Excellence ORIGINS (EXC-2094)---390783311 is acknowledged. Computer  resources  for  this  project  have been  provided  by  the  Leibniz  Supercomputing   Centre (LRZ)  under  LRZ  project  ID  pr53yi  and  by  the  Max Planck Computing and Data Facility (MPCDF) on the HPC system Draco.

\begin{appendix}
\section{Comparison of results with  linear and cubic polynomials for $F(\gamma)$}

\begin{figure*}
  \centering
\includegraphics[width=0.45\textwidth]{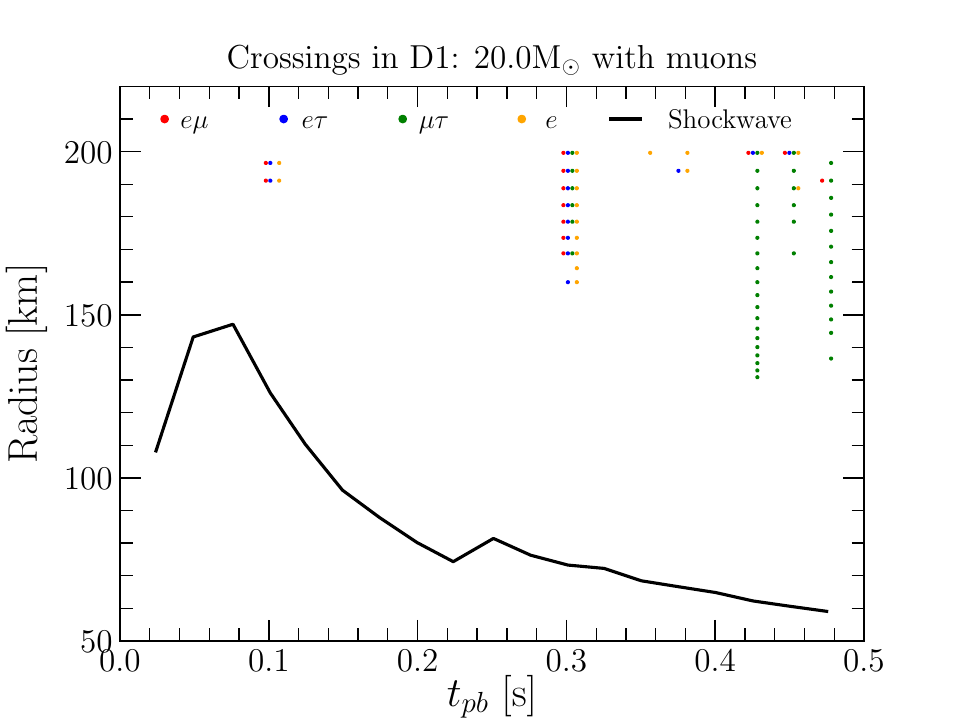}
 \includegraphics[width=0.45\textwidth]{SFHo_20_Msol_with_muons_summary.pdf}
 \includegraphics[width=0.45\textwidth]{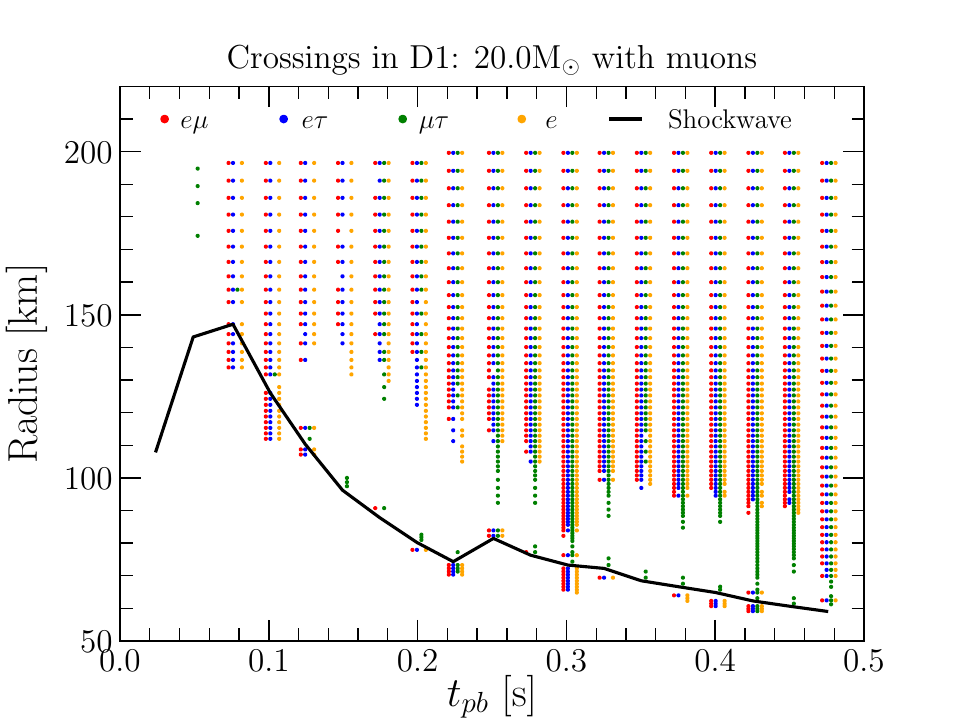}
       \caption{Each panel has the same format as Fig.~\ref{fig:summary_plot_D1}. The top left panel is obtained using linear polynomials for $F(\gamma)$, the top right one with quadratic polynomials (same as in the figures in the main text), the bottom one with cubic polynomials.}
        \label{fig:polynomials_comparison}
\end{figure*}

Figure~\ref{fig:polynomials_comparison} compares the pattern crossings obtained with three types of polynomials: linear (left panel), quadratic (central panel) and cubic (right panel). Here we just refer to the model for a 20$M_\odot$ progenitor with muons in D1. In the linear case, very few crossings are found. On the other hand, using cubic polynomials does not change the conclusions listed in the main text, since the crossing distribution is very similar to the one obtained with quadratic polynomials. We consider the choice of a linear functional form for $F(\gamma)$ as too pessimistic.

\begin{figure*}[h]
  \centering
  \includegraphics[width=0.45\textwidth]{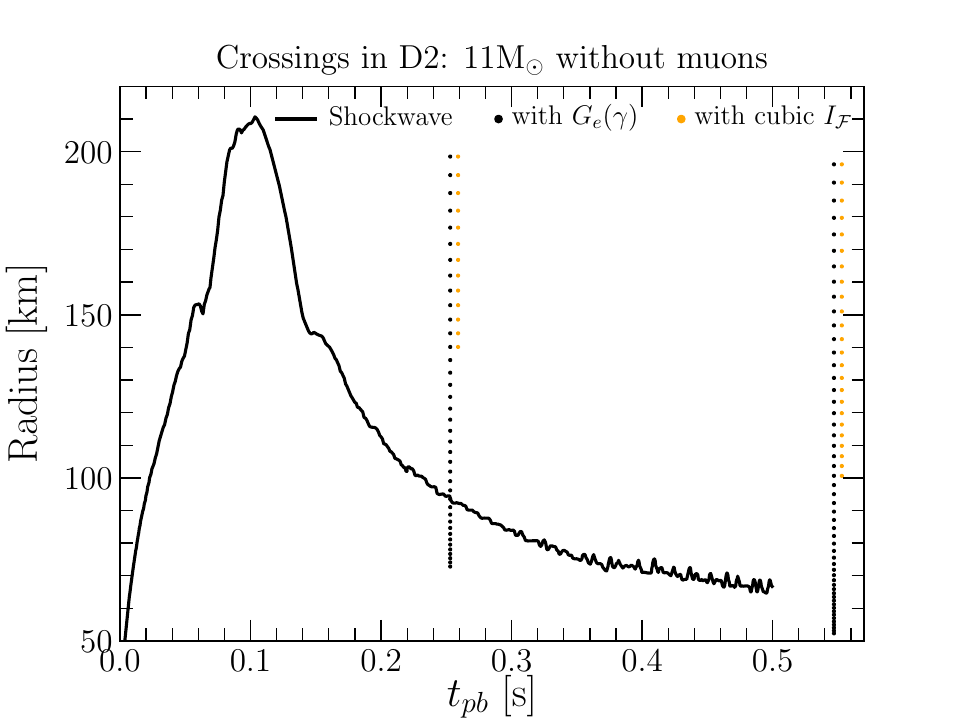}
    \includegraphics[width=0.45\textwidth]{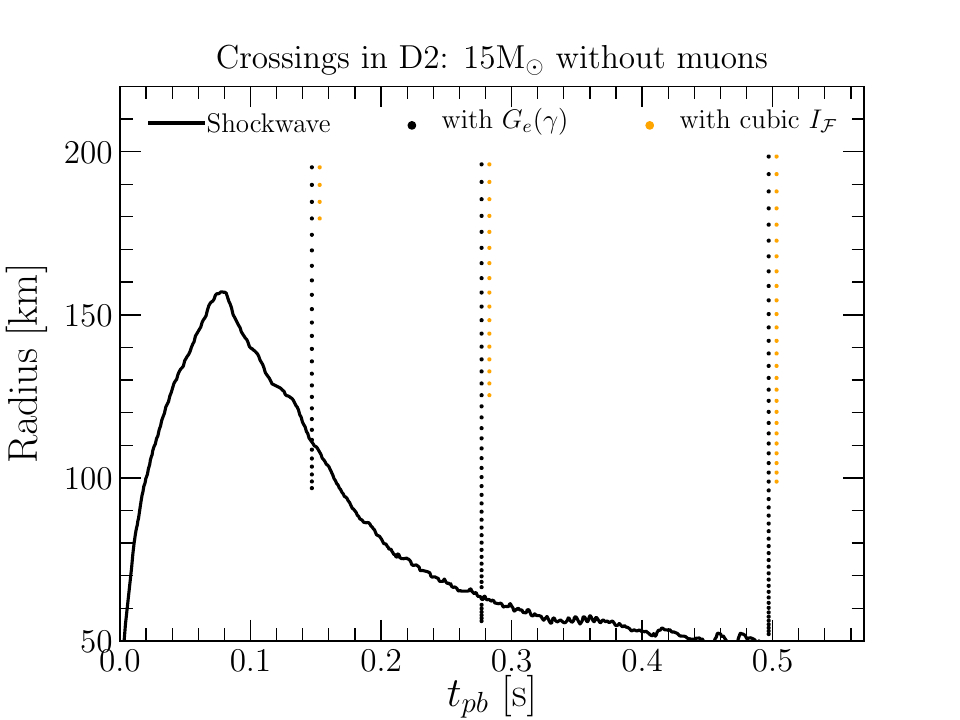}
    \includegraphics[width=0.45\textwidth]{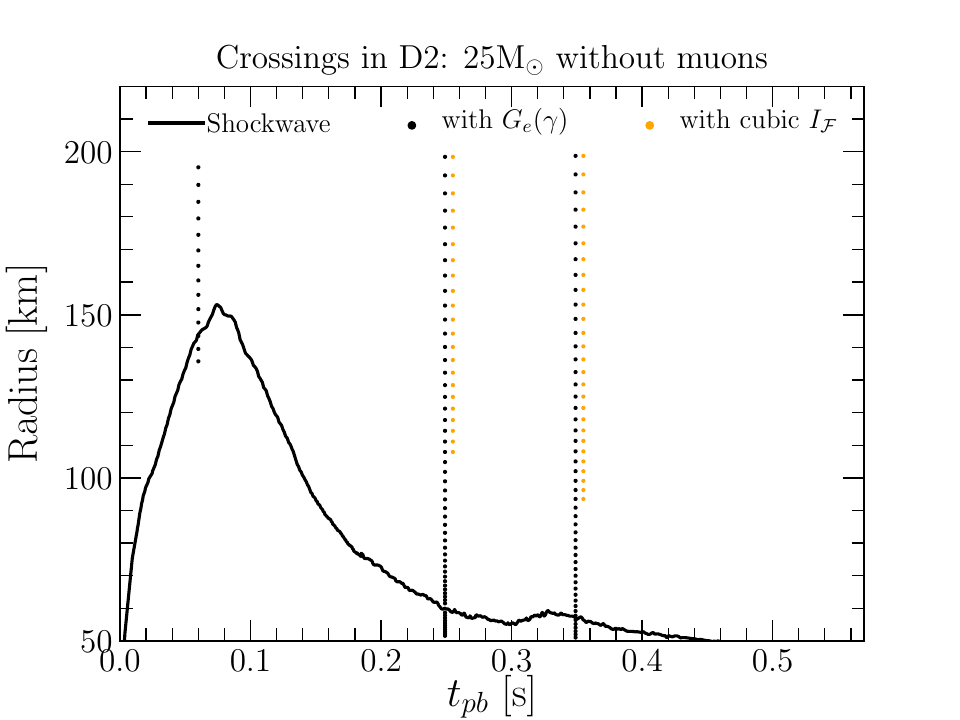}
       \caption{Same as Fig. \ref{fig:summary_plot_D2} but with the addition of crossings found with the moments method using cubic polynomials (yellow dots).}
        \label{fig:summary_plot_D2_with_cubic_polynomials}
\end{figure*}

\begin{figure*}[h]
  \centering
  \includegraphics[width=0.45\textwidth]{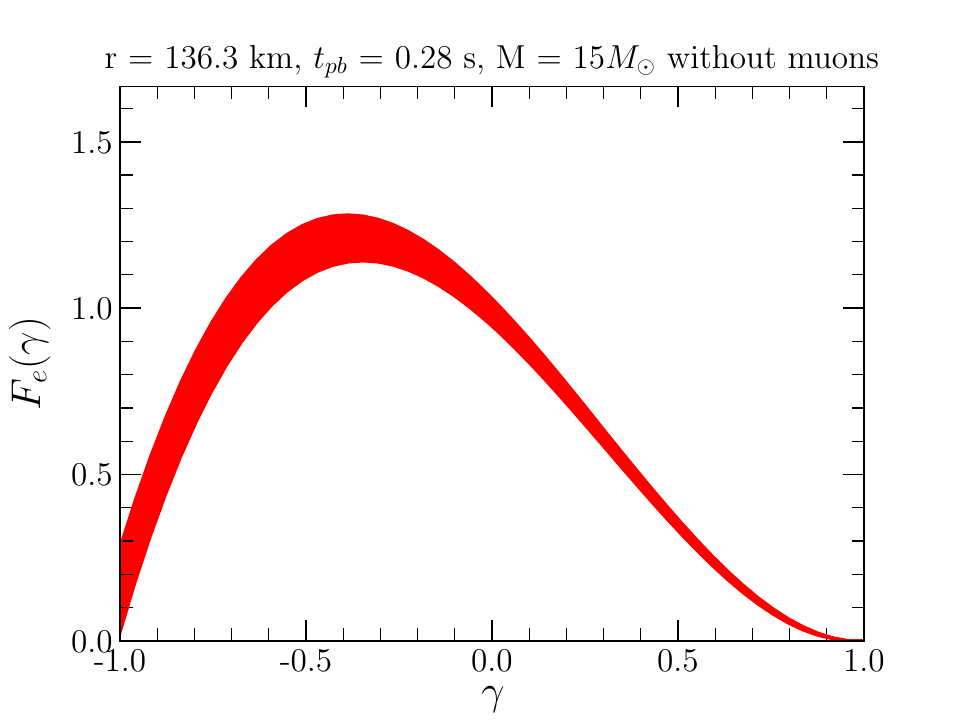}
       \caption{Shape of \ $\mathcal{F}_{e\mu}(\gamma)$ for which crossings are found at $t_{pb}=0.28$ s  and $r=136.3$ km. Here we have considered the 15$M_\odot$ model without muons of D2. $\mathcal{F}(\gamma)$ is assumed to be a cubic polynomial, because with a quadratic one no crossing are found.}
        \label{fig:F_shape_D2}
\end{figure*}

\section{Analysis of D2 with cubic polynomials}

Fig.~\ref{fig:summary_plot_D2_with_cubic_polynomials} is equivalent to Fig.~\ref{fig:summary_plot_D2} in the main text, but it displays also the crossing points we find assuming cubic polynomials. We emphasize again that no crossings are found in the radial range $r<200$ km of D2 when using quadratic polynomials.

Fig.~\ref{fig:F_shape_D2} shows an example of shape of $\mathcal{F}(\gamma)$ that allows the crossings in Fig.~\ref{fig:summary_plot_D2_with_cubic_polynomials} to be found. Such a shape is different from the one obtained for D1 and shown in Fig. \ref{fig:F_shape}. Indeed, in D2 $\mathcal{F}$ is vanishing at both $\gamma=\pm1$, whereas in D1 this happens only at $\gamma+1$. The shape is D2 seems to be compatible with the double crossings observed in the angular distributions shown in Fig.~\ref{fig:angular_distributions}.

\end{appendix}

\bibliographystyle{JHEP}
\bibliography{Biblio}

\end{document}